# Portfolio Optimization in the Financial Market with Correlated Returns under Constraints, Transaction Costs and Different Rates for Borrowing and Lending

VLADIMIR DOMBROVSKII, TATYANA OBEDKO
Department of Economics
Tomsk State University
Lenina Street 36, Tomsk
RUSSIAN FEDERATION
dombrovs@ef.tsu.ru, tatyana.obedko@mail.ru

*Abstract:* - In this work, we consider the optimal portfolio selection problem under hard constraints on trading amounts, transaction costs and different rates for borrowing and lending when the risky asset returns are serially correlated. No assumptions about the correlation structure between different time points or about the distribution of the asset returns are needed. The problem is stated as a dynamic tracking problem of a reference portfolio with desired return. Our approach is tested on a set of a real data from Russian Stock Exchange MICEX.

## 1 Introduction

The investment portfolio (IP) management is an area of both theoretical interest and practical importance. The foundation for modern portfolio selection theory is the single-period mean-variance approach suggested by Markowitz [13] and the Merton's [14] IP model in continuous time. At present, there exists a variety of models and approaches to the solution of the IP optimization problem, but most of them are the complications and extensions of the Markowitz and Merton approaches to various versions of stochastic models of the prices of risky and risk-free assets and utility functions. Most studies assume the time independency of the return vector, with a few exceptions.

The vast majority of the existing literature on dynamic portfolio selection is based on using dynamic programming approach for determining the solution. However, that approach leads to the well-known "curse of dimensionality," which hinders design of the decision strategies under constraints. Therefore, the most of the results presented in the literature are limited to the cases without trading constraints and transaction costs. Also, the rates of borrowing and lending are assumed to be the same. However, it's well-known that realistic investment models must include these features [11].

In this paper, we consider the dynamic investment portfolio selection problem subject to hard constraints on trading amounts (a borrowing limit on the total wealth invested in the risky assets, and long- and shortsale restrictions on all risky assets), taking into account the presence of quadratic transaction costs. Other realistic feature we incorporate is that in our model the rates of borrowing and lending are different (the rate of borrowing is greater than that of lending).

Empirical evidence shows that the returns of the risky assets always exhibit certain degree of dependency among time periods, e.g., see [8,19] and reference therein. We assume also that the risky asset returns are serially correlated. The only conditions imposed on the distributions of the asset returns are the existences of the conditional mean vectors and of the conditional second-order moments. No assumptions about the correlation structure between different time points or about the distribution of the asset returns are needed.

The problem is stated as a dynamic tracking problem of a reference portfolio with desired return.

The investor's objective is to choose the dynamic trading strategy to minimize the conditional mean-square error between the investment portfolio value and a reference (benchmark) portfolio, penalized for the transaction costs assosiated with trading. We consider quadratic transaction costs. The natural interpretation of a quadratic cost is that price impact is linear in the trade size, resulting in a quadratic cost [15].

In this work, we use the model predictive control (also known as receding horizon control) method in order to solve the problem. The major attraction of such technique lies in the fact that it can handle hard constraints on the inputs (manipulated variables) and states/outputs of a process and allows to avoid the "curse of dimensionality" [6-8,10,18-20].

There are many examples of the MPC in finance applications. Some recent works can be found in [1,4-7,9,16,17]. In [4-7,9] investment portfolio optimization with constraints using MPC is considered. Dynamic option hedging using MPC is presented in [16] and in [1]. In all of these papers, authors assume the hypothesis of serially independent returns and/or consider the explicit form of the model describing the price process of the risky assets (e.g., geometric Brownian motion, etc.). Related results in multi-period portfolio optimization can be found in [2-3] where a multi-stage optimization model is developed. In a developed model portfolio, diversity constraints are imposed in expectation (soft constraints). Calafiore [2-3] proposed an approximated technique to solve the problem via stochastic simulations of the return series that can be used in practice when a full stochastic model for return dynamics is available.

The purpose of the present paper is to provide numerically tractable algorithm for practical applications. We want to demonstrate the performance of our model under real market conditions. We pay a particular attention to testing of our approach on a set of a real data from the Russian Stock Exchange MICEX.

This work is organized as follows. Section 2 presents portfolio model and the optimization problem formulation. The main results of this article are presented in Section 3 where we design the optimal investment strategy for the problem under consideration. In Section 4 the numerical modeling results are presented. This paper is concluded in Section 5 with some final remarks.

## 2 Portfolio Model and Optimization Problem

### 2.1 The Proposed Portfolio Model

Let us consider the investment portfolio of $n$ risky assets and one risk-free asset (e.g. a bank account or a government bond). Let $u_i(k)$, $(i=0,1,2,...,n)$ denote the amount of money invested in the $i$th asset at time $k$; $u_0(k) \geq 0$ is the amount invested in a risk-free asset. Investor also can borrow the capital in case of need. The volume of the borrowing of the risk-free asset is equal to $u_{n+1}(k) \geq 0$. If $u_i(k)<0$, $(i=1,2,...,n)$, then we use short position with the amount of shorting $|u_i(k)|$. The wealth process $V(k)$ satisfies

$$V(k) = \sum_{i=1}^{n} u_i(k) + u_0(k) - u_{n+1}(k). \quad (1)$$

Let $P_i(k)$ denote the market value of the $i$th risky asset at time $k$, and $\eta_i(k+1)$ denote the corresponding return per period $[k,k+1]$, defined as

$$\eta_i(k+1) = \frac{P_i(k+1) - P_i(k)}{P_i(k)}.$$

It is a stochastic value unobserved at time $k$.

We consider self-financing portfolio. Self-financing means that the whole wealth obtained at the trading period $k$ will be exactly reinvested at the trading period $k+1$. By considering the self-finance strategies, the wealth dynamics are given by

$$V(k+1) = \sum_{i=1}^{n}\left[1+\eta_i(k+1)\right]u_i(k) + \left[1+r_1\right]u_0(k) - \left[1+r_2\right]u_{n+1}(k), \quad (2)$$

with initial value $V(0)$, where $r_1$ is the riskless lending rate, $r_2$ is the riskless borrowing rate ($r_1<r_2$).

Using (1), the dynamics (2) can be rewritten as follows

$$V(k+1) = \left[1+r_1\right]V(k) + \sum_{i=1}^{n}\left[\eta_i(k+1)-r_1\right]u_i(k) - \left[r_2-r_1\right]u_{n+1}(k), \quad (3)$$

here $u_0(k) = V(k) - \sum_{i=1}^{n} u_i(k) + u_{n+1}(k)$ is the amount invested in a risk-free asset.

We impose the following constraints on the decision variables (a borrowing limit on the total wealth invested in the risky assets, and long- and short-sale restrictions on all risky assets)

$$u_i^{\min}(k) \leq u_i(k) \leq u_i^{\max}(k), (i=\overline{1,n}), \quad (4)$$

$$0 \leq V(k) - \sum_{i=1}^{n} u_i(k) + u_{n+1}(k) \leq u_0^{\max}(k), \quad (5)$$

$$0 \leq u_{n+1}(k) \leq u_{n+1}^{\max}(k). \quad (6)$$

If $u_i^{\min}(k)<0$, $(i=1,2,...,n)$, so we suppose that the amounts of the short-sale are restricted by $|u_i^{\min}(k)|$; if the short-selling is prohibited then $u_i^{\min}(k) \geq 0$, $(i=1,2,...,n)$. The amounts of long-sale are restricted by $u_i^{\max}(k)$, $(i=1,2,...,n)$; $u_0^{\max}(k) \geq 0$ defined the maximum amount of money we can invest in the risk-free asset; the borrowing amount is restricted by $u_{n+1}^{\max}(k) \geq 0$. Note, that values $u_i^{\min}(k)$, $(i=0,1,...,n)$, $u_i^{\max}(k)$, $(i=0,1,...,n+1)$ are often depend on common wealth of portfolio in practice. So that we can write $u_i^{\min}(k)=\beta_i V(k)$, $u_i^{\max}(k)=\gamma_i V(k)$, where $\beta_i$, $\gamma_i$ are constant parameters.

Let $\mathbb{F}=(\mathfrak{F}_k)_{k \geq 1}$ be the complete filtration with $\sigma$-field $\mathfrak{F}_k$ generated by the $\{\eta(s): s=0, 1, 2,...,k\}$ that models the flow of information about asset returns to time $k$.

Let us assume that the vectors of risky asset returns $\eta(k)=[\eta_1(k)\ \eta_2(k)\ \ldots\ \eta_n(k)]^T$, $k=0,1,\ldots$, form a serially correlated non-stationary discrete-time multivariate process with finite conditional moments

$$E\{\eta(k+i)/\mathfrak{F}_k\} = \overline{\eta}(k+i),$$
$$E\{\eta(k+i)\eta^T(k+j)/\mathfrak{F}_k\} = \Theta_{ij}(k),$$
$$(i,j=\overline{1,l}; k=0,1,2,\ldots).$$

Therefore, the lead-lag relationships between component series $\eta_t(k+i)$ and $\eta_f(k+j)$ are described by the matrices $\Theta_{ij}(k)$ of the second-order conditional moments.

Throughout the paper, we use the following notations. For any matrix $\psi[\eta(k+i),k+i]$, dependent on $\eta(k+i)$, $\overline{\psi}(k+i)=E\{\psi[\eta(k+i),k+i]/\mathfrak{F}_k\}$, without indicating the explicit dependence of matrices on $\eta(k+i)$.

One motivation for such a model is the fact that a large number of empirical analyses of assets' price dynamics show that there exists salient serial correlations in the returns of financial assets [8,19].

## 2.2 Optimization Problem (Risk Function)

Our objective is to control the investment portfolio, via dynamics asset allocation among the $n$ stocks and the risk-free asset, as closely as possible tracking the deterministic benchmark

$$V^0(k+1) = [1+\mu_0]V^0(k), \qquad (7)$$

where $\mu_0$ is a given parameter representing the growth factor, the initial state is $V^0(0)=V(0)$.

We use the MPC methodology in order to define the optimal control portfolio strategy. For the given prediction horizon $m$, a sequence of predictive controls (trading amounts) $u(k/k)$, $u(k+1/k)$,…, $u(k+m-1/k)$ depending on the portfolio wealth at the current time $k$ and all the information about asset returns to time $k$ is calculated at each step $k$. This sequence optimizes the criterion chosen by the investor for the prediction horizon. At the time $k$, $u(k)=u(k/k)$ is assumed to be control $u(k)$. To obtain the control at the next step $k+1$, the procedure is repeated, and the control horizon is one step shifted.

We consider the following objective with receding horizon (risk function)

$$J(k+m/k) = E\left\{\sum_{i=1}^m \left[V(k+i)-V^0(k+i)\right]^2 - \right.$$
$$-\rho(k,i)\left[V(k+i)-V^0(k+i)\right]/V(k),\mathfrak{F}_k\right\} + \qquad (8)$$
$$+\sum_{i=0}^{m-1} E\left\{\left[u(k+i/k)-u(k+i-1/k)\right]^T R(k,i) \times \right.$$
$$\left. \times \left[u(k+i/k)-u(k+i-1/k)\right]/V(k),\mathfrak{F}_k\right\},$$

where $m$ is the prediction horizon, $u(k+i/k)=[u_1(k+i/k),\ldots,u_{n+1}(k+i/k)]^T$ is the predictive control vector, $i=\overline{0,m-1}, k=0,1,2,\ldots;$ $u(k-1/k)=u(k-1)$ is the optimal control vector obtained on the previous step, $u(-1/0)=0$; $R(k,i)>0$ is a positive-definite symmetric matrix mesuring the level of transaction costs, $\rho(k,i)>0$ is a positive weight coefficient; $E\{a/b\}$ is the conditional expectation of $a$ with respect to $b$. Notice that variable $V^0(k)$ is known for all time instant $k$ and may be considered as a pre-chosen parameter.

Let us explain the terms in the objective function (8). The first term represents the conditional mean-square error between the investment portfolio value and a reference (benchmark) portfolio, the second term penalizes wealth values that less than the desired value. The third term penalizes for transaction costs assosiated with trading amount $|u(k+i/k)-u(k+i-1/k)|$.

An important advantage of tracking a reference portfolio approach under quadratic criterion (8) is its capability to predict the trajectory of growth portfolio wealth, which would follow close to the deterministic (given by the investor) benchmark or beat it. It makes possible to obtain a smooth curve of the growth of the portfolio wealth on the entire investment horizon. It is one of the basic requirements for the trading strategies of investors in financial markets. The growth factor $\mu_0$ is selected by investor, based on the analysis of the financial market.

## 3 The Proposed Investment Strategy Design

The problem of minimizing the criterion (8) is equivalent to the quadratic control problem with criterion

$$J(k+m/k) = E\{\sum_{i=1}^m V^2(k+i) - \qquad (9)$$
$$-R_1(k,i)V(k+i)/V(k),\mathfrak{F}_k\} +$$
$$+\sum_{i=0}^{m-1} E\left\{\left[u(k+i/k)-u(k+i-1/k)\right]^T R(k,i) \times \right.$$
$$\left. \times \left[u(k+i/k)-u(k+i-1/k)\right]/V(k),\mathfrak{F}_k\right\},$$

where we eliminated the term that is independent of control variables, $R_1(k+i)=2V^0(k+i)+\rho(k,i)$.

We have the following theorem.

**Theorem 1.** *Let the wealth dynamics is given by (3) under constraints (4)-(6). Then the MPC policy with receding horizon m, such that it minimizes the objective (9), for each instant k is defined by the equation*

$$u(k) = [I_{n+1} \quad 0_{n+1} \quad ... \quad 0_{n+1}]U(k), \quad (10)$$

*where $I_{n+1}$ is (n+1)-dimensional identity matrix; $0_{n+1}$ is (n+1)-dimensional zero matrix; $U(k)=[u^T(k/k),...,u^T(k+m-1/k)]^T$ is the set of predictive controls defined from the solving of quadratic programming problem with criterion*

$$Y(k+m/k) = [2V(k)G(k) - F(k)]U(k) + \\ + U^T(k)[H(k) + \overline{R}(k)]U(k) \quad (11)$$

*under constraints (element-wise inequality)*

$$U_{\min}(k) \leq \overline{S}U(k) \leq U_{\max}(k), \quad (12)$$

*where*

$$U_{\min}(k) = [u^T_{\min}(k), 0_{n+2\times 1}, ..., 0_{n+2\times 1}]^T,$$
$$U_{\max}(k) = [u^T_{\max}(k), 0_{n+2\times 1}, ..., 0_{n+2\times 1}]^T,$$

$$u_{\min}(k) = \begin{bmatrix} u_1^{\min}(k) \\ u_2^{\min}(k) \\ ... \\ u_n^{\min}(k) \\ -V(k) \\ 0 \end{bmatrix}, u_{\max}(k) = \begin{bmatrix} u_1^{\max}(k) \\ u_2^{\max}(k) \\ ... \\ u_n^{\max}(k) \\ u_0^{\max}(k) - V(k) \\ u_{n+1}^{\max}(k) \end{bmatrix},$$

$$\overline{S} = \text{diag}\{S, 0_{n+2\times n+1}, ..., 0_{n+2\times n+1}\},$$

$$S = \begin{bmatrix} 1 & 0 & ... & 0 & 0 \\ 0 & 1 & ... & 0 & 0 \\ ... & ... & ... & ... & ... \\ 0 & 0 & ... & 1 & 0 \\ -1 & -1 & ... & -1 & 1 \\ 0 & 0 & ... & 0 & 1 \end{bmatrix},$$

$$\overline{R}(k) = \begin{bmatrix} R(k,0)+R(k,1) & -R(k,1) & ... \\ -R(k,1) & R(k,1)+R(k,2) & ... \\ ... & ... & ... \\ 0_{n+1\times n+1} & 0_{n+1\times n+1} & ... \\ 0_{n+1\times n+1} & 0_{n+1\times n+1} & ... \\ \\ ... & 0_{n+1\times n+1} & 0_{n+1\times n+1} \\ ... & 0_{n+1\times n+1} & 0_{n+1\times n+1} \\ ... & ... & ... \\ ... & R(k,m-1)+R(k,m) & -R(k,m) \\ ... & -R(k,m) & R(k,m) \end{bmatrix},$$

*H(k), G(k), F(k) are the block matrices*

$$H(k) = [H_{tf}(k)], G(k) = [G_t(k)], F(k) = [F_t(k)], \\ (t,f=\overline{1,m}), \quad (13)$$

*and the blocks satisfy the following recursive equations*

$$H_{tt}(k) = E\{b^T[\eta(k+t), k+t] \times \\ \times Q_1(m-t)b[\eta(k+t), k+t]/\mathfrak{F}_k\}, \quad (14)$$

$$H_{tf}(k) = A^{f-t}E\{b^T[\eta(k+t), k+t] \times \\ \times Q_1(m-f)b[\eta(k+f), k+f]/\mathfrak{F}_k\}, t<f, \quad (15)$$

$$H_{tf}(k) = H^T_{ft}(k), t>f, \quad (16)$$

$$G_t(k) = A^t Q_1(m-t)\overline{b}(k+t), \quad (17)$$

$$F_t(k) = Q_2(m-t)\overline{b}(k+t) - 2R(k,0)u(k-1), \quad (18)$$

$$Q_1(t) = A^2 Q_1(t-1) + 1, Q_1(0) = 1,$$
$$Q_2(t) = AQ_2(t-1) + R_1(k,m-t), Q_2(0) = R_1(k,m),$$
$$R_1(k,t) = 2V^0(k+t) + \rho(k,t), (t=\overline{1,m}),$$
$$A = 1 + r_1,$$
$$b[\eta(k), k] = [\eta_1(k) - r_1 \quad ... \quad \eta_n(k) - r_1 \quad r_1 - r_2].$$

A brief proof of this theorem is reported in the Appendix.

## 3 A real data numerical example

In this section, we present several numerical examples demonstrating the application of our approach to portfolio of a real stocks. We want to assess the performance of our model under real market conditions by computing the portfolio wealth over a long period of time. The data used for these examples are taken from the Russian Stock Exchange MICEX (www.finam.ru). They include the daily stock prices of the largest Russian companies such as Sberbank, Gazprom, VTB, LUKOIL, NorNickel, Rosneft, and Sibneft. The portfolio was composed of five risky assets. Performing numerical modelling, we looked over all of the possible combinations of the five assets.

We consider the situation of an investor who has to allocate one unit of wealth over the investment horizon of approximately 1500 trading days (about four years) among risky assets and one risk-free asset. The risk-free asset is considered here as a bank account with $r_1$=0.0001, $r_2$=0.0002. The updating of the portfolio based on the MPC is executed once every trading day.

We set the tracking target to return 0.15% per day ($\mu_0$=0.0015). We assumed an initial portfolio wealth of $V(0)=V^0(0)=1$. The matrix measuring the level of transaction costs is set as $R(k,i)=\text{diag}(10^{-4},...,10^{-4})$ for all $k,i$, the weight coefficient $\rho(k,i)=0.1$ for all $k,i$. We impose constraints on the tracking portfolio problem with parameters $\beta_i$=-0.6, ($i$=1,...,$n$), $\gamma_i$=3, ($i$=1,...,$n+1$). Therefore we allow borrowing and short selling.

For the on-line finite horizon MPC problems, we used a horizon of $m$=10, and numerically solved it in MATLAB by using the quadprog.m function.

At each time *k*, the optimization problem requires as input parameters the predicted returns and predicted second moments of returns over the predictive horizon *m*. These parameters can be estimated using different model specifications describing the return asset evolution. Examples include using autoregressive models, conditional heteroscedastic models, factor models, complex nonparametric methods and others (see, for instance, [12,19]).

As a simple example, we assume that the multivariate process of risky asset returns follows the VAR(2) model (vector autoregressive model of order 2) [12]

$$\eta(k+1) = \nu + A_1\eta(k) + A_2\eta(k-1) + \omega(k+1),$$

where $A_1, A_2$ are the coefficient matrices,
$\nu = (I_n - A_1 - A_2)\mu$ is a vector of intercept terms,
$\mu = E\{\eta(k)\}$; and $\omega(k+1)$ is an *n*-dimensional white noise, that is,

$$E\{\omega(k+1)\} = 0;\ E\{\omega(k+1)\omega^T(k+1)\} = \sigma;$$
$$E\{\omega(k+i)\omega^T(k+j)\} = 0, i \neq j.$$

The covariance matrix $\sigma$ is assumed to be nonsingular.

We estimated parameters of this model by the ordinary least squares method using the observed historical data based on the past 200 trading days prior to the tracking period. These parameters were considered constant along the entire period under study and equal to the initial empirical estimates, based on backwards data. We calculated the predicted conditional second moments based on this VAR(2) model and substituted them into equations (14)-(15).

In practice, time series of risky asset returns have a trending behaviour which is not compatible with the assumptions of the classical VAR model. In order to capture short-run trends of risky asset returns, we use the following modification of the forecasting procedure based on the VAR(2) model. We calculate the sample means of returns $\hat{\eta}(k)$ using 2-day windows of past historical return data and incorporate these estimates in the VAR(2) - predictor

$$\hat{E}\{\eta(k+h)/\eta(k)\} = \hat{\nu} + \hat{A}_1\hat{E}\{\eta(k+h-1)/\eta(k)\} + $$
$$+ \hat{A}_2\hat{E}\{\eta(k+h-2)/\eta(k)\},$$

where the true coefficients $\nu, A_1$ are replaced by estimators

$$\hat{\nu}, \hat{A}_1; \hat{\nu} = (I_n - \hat{A}_1 - \hat{A}_2)\hat{\eta}(k); h = 1, 2, \ldots, m.$$

This formula was used for recursively computing the *h*-step predictors starting with *h*=1.

This predictor is used to predict the expected returns over the predictive horizon *m* at each decision time *k* in equations (17) and (18). When a new measurement becomes available, the oldest measurement is discarded and the new measurement is added. So, we use the adjusted procedure, updating the estimates of mean returns at each time *k*.

One motivation for such a heuristic approach is that we have no restrictions to construct any type of predictors in order to obtain the best asset allocation strategies. However, forecasting is too large a topic to address adequately in this work and the investigation of the sensitivity of optimization results to the estimated parameters is outside the scope of this work.

We present the typical results of the experiments on fig. 1-3. In the pictures below, the portfolio was composed of five risky assets: LUKOIL, Gazprom, Sberbank, Rosneftj, and NorNickel. Investment period is from 20.07.2007 to 11.09.2014 (approximately 6 years). Fig. 1 plots the tracking portfolio and a reference portfolio values. In fig. 2, we have investments in the risky asset Gazprom. Fig. 3 plots risky asset returns for asset Gazprom.

Several insights can be gathered from the examples illustrated above.

Fig. 1 shows that the tracking a reference portfolio strategy allows us to obtain a smooth curve of growth. The advantage of the control according to the quadratic criterion is that it is possible to predict the trajectory of the growth of portfolio wealth, which should follow as close as possible to the deterministic benchmark given by the investor.

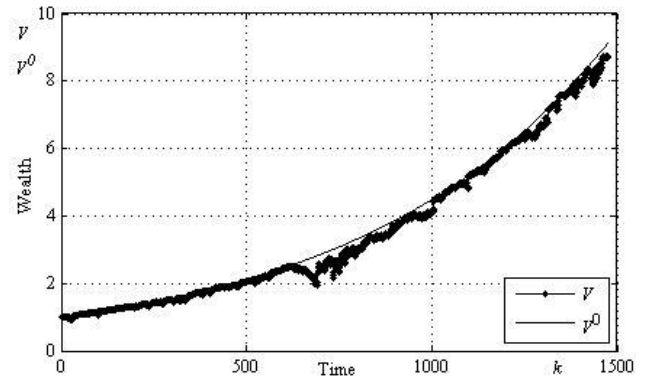

Fig.1. Tracking performance (*V* – real portfolio, $V^0$ – reference portfolio).

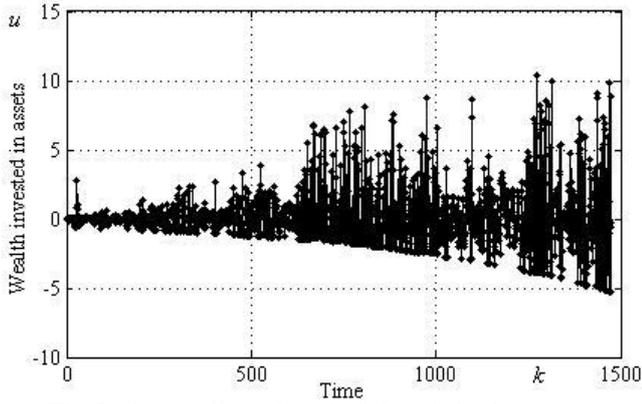

Fig.2. Asset allocation decision ($u$ is the amount invested in Gazprom).

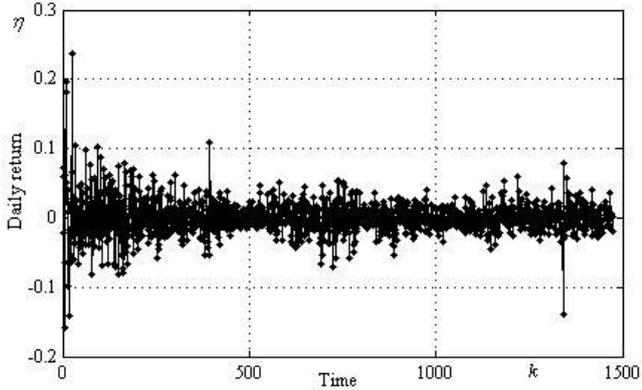

Fig.3. Risky asset returns (Gazprom).

It is important to acknowledge that in our experiments, where we use a rather simple model for parameters estimation, the performance of proposed strategies appears to be rather efficient. So, our approach allows us to design strategies which are desensitized, i.e., robustified, to parameters estimation. It is clear that one can use more sophisticated estimation schemes to improve the precision of parameters estimation.

## 4 Conclusion and future work

In this paper, we studied a discrete-time portfolio selection problem with serially correlated returns, for which only the first and the second conditional moments are known. The knowledge of the statistical distributions of the returns is not assumed. We proposed to use the MPC methodology in order to solve the problem. The optimal portfolio control strategy was derived under hard constraints on trading amounts, transaction costs and different rates for borrowing and lending. The advantage of using a receding horizon implementation is that at each decision stage we can profit from observations of actual market behavior during the preceding period and use information to feed fresh estimates to the model.

We presented the numerical modeling results, based on a set of real data from the Russian Stock Exchange MICEX. We find that on actual data the proposed approach is reasonable. The value of the portfolio follows the value of the reverence portfolio, beating it most of the time and the constraints are satisfied.

## Appendix

**Proof of the Theorem 1:** The portfolio dynamics (3) can be rewritten in the form

$$V(k+1) = [1+r_1]V(k) + b[\eta(k+1), k+1]u(k),$$

where $\eta(k) = [\eta_1(k)\ \eta_2(k)\ \ldots\ \eta_n(k)]^T$ is the vector of risky asset returns, $u(k) = [u_1(k)\ u_2(k)\ \ldots\ u_{n+1}(k)]^T$ is the vector of input (manipulated) variables, and

$$b[\eta(k), k] = [\eta_1(k) - r_1\ \ldots\ \eta_n(k) - r_1\ r_1 - r_2].$$

Constraints (4)-(6) can be rewritten in matrix form (element-wise inequality):

$$u_{\min}(k) \leq Su(k) \leq u_{\max}(k), \quad (19)$$

where

$$S = \begin{bmatrix} I_n & 0_{n \times 1} \\ -E & 1 \\ 0_{1 \times n} & 1 \end{bmatrix}, E = [1\ \ldots\ 1]_{1 \times n},$$

$$u_{\min}(k) = \begin{bmatrix} u_1^{\min}(k) \\ u_2^{\min}(k) \\ \ldots \\ u_n^{\min}(k) \\ -V(k) \\ 0 \end{bmatrix}, u_{\max}(k) = \begin{bmatrix} u_1^{\max}(k) \\ u_2^{\max}(k) \\ \ldots \\ u_n^{\max}(k) \\ u_0^{\max}(k) - V(k) \\ u_{n+1}^{\max}(k) \end{bmatrix}.$$

The objective (9) can be written in the form

$$J(k+m/k) = E\{X^T(k+1)X(k+1) - \quad (20)$$
$$-\Delta_1(k+1)X(k+1) + U^T(k)\overline{R}(k)U(k) -$$
$$-2u(k/k)R(k,0)u(k-1) +$$
$$+u^T(k-1)R(k,0)u(k-1)/V(k), \mathfrak{F}_k\},$$

subject to

$$X(k+1) = \Psi V(k) + \Phi[\Xi(k+1), k+1]U(k), \quad (21)$$

where

$$X(k+1) = \begin{bmatrix} V(k+1) \\ V(k+2) \\ \ldots \\ V(k+m) \end{bmatrix}, \Psi = \begin{bmatrix} A \\ A^2 \\ \ldots \\ A^m \end{bmatrix}, A = 1+r_1,$$

$$U(k) = [u^T(k/k), u^T(k+1/k), \ldots, u^T(k+m-1/k)]^T,$$

$$\Xi(k+1) = \begin{bmatrix} \eta(k+1) \\ \eta(k+2) \\ \ldots \\ \eta(k+m) \end{bmatrix},$$

$$\Phi[\Xi(k+1),k+1] =$$

$$= \begin{bmatrix} b[\eta(k+1),k+1] & 0_{1\times n+1} & \cdots \\ Ab[\eta(k+1),k+1] & b[\eta(k+2),k+2] & \cdots \\ \cdots & \cdots & \cdots \\ A^{m-1}b[\eta(k+1),k+1] & A^{m-2}b[\eta(k+2),k+2] & \cdots \end{bmatrix}$$

$$\begin{matrix} \cdots & 0_{1\times n+1} \\ \cdots & 0_{1\times n+1} \\ \cdots & \cdots \\ \cdots & b[\eta(k+m),k+m] \end{matrix},$$

$$\Delta_1(k+1) = [R_1(k,1), R_1(k,2), \ldots, R_1(k,m)],$$

$$\bar{R}(k) = \begin{bmatrix} R(k,0)+R(k,1) & -R(k,1) & \cdots \\ -R(k,1) & R(k,1)+R(k,2) & \cdots \\ \cdots & \cdots & \cdots \\ 0_{n+1\times n+1} & 0_{n+1\times n+1} & \cdots \\ 0_{n+1\times n+1} & 0_{n+1\times n+1} & \cdots \end{bmatrix}$$

$$\begin{matrix} \cdots & 0_{n+1\times n+1} & 0_{n+1\times n+1} \\ \cdots & 0_{n+1\times n+1} & 0_{n+1\times n+1} \\ \cdots & \cdots & \cdots \\ \cdots & R(k,m-1)+R(k,m) & -R(k,m) \\ \cdots & -R(k,m) & R(k,m) \end{matrix}.$$

Using (21), we can rewrite (20) as follows

$$J(k+m/k) = V^2(k)\Psi^T\Psi - \Delta_1\Psi V(k) + \\ + \left\{ \left[ 2V(k)\Psi^T - \Delta_1(k+1) \right] \times \\ \times E\{\Phi[\Xi(k+1),k+1]/\mathfrak{F}_k\} + L(k) \right\} U(k) + \\ + U^T(k)E\{\Phi^T[\Xi(k+1),k+1] \times \\ \times \Phi[\Xi(k+1),k+1]/\mathfrak{F}_k\} U(k) + \\ + u^T(k-1)R(k,0)u(k-1), \quad (22)$$

where

$$L(k) = [2R(k,0)u(k-1) \quad 0_{1\times n+1} \quad \cdots \quad 0_{1\times n+1}].$$

Denote the matrices

$$H(k) = E\{\Phi^T[\Xi(k+1),k+1]\Phi[\Xi(k+1),k+1]/\mathfrak{F}_k\},$$

$$G(k) = \Psi^T E\{\Phi[\Xi(k+1),k+1]/\mathfrak{F}_k\},$$

$$F(k) = \Delta_1(k+1)E\{\Phi[\Xi(k+1),k+1]/\mathfrak{F}_k\} + L(k).$$

We have that the minimization of the criterion (9) under constraints (4)-(6) is equivalent to the quadratic programming problem with criterion

$$Y(k+m/k) = [2V(k)G(k) - F(k)]U(k) + \\ + U^T(k)[H(k) + \bar{R}(k)]U(k)$$

under constraints (19).

Straightforward calculations lead to the expressions (14)-(18) for the matrices $H(k)$, $G(k)$, $F(k)$. This completes the proof.